# Electronic conduction in a three-terminal molecular transistor

Haiying He and Ravindra Pandey[*]
Department of Physics and Multi-scale Technology Institute,
Michigan Technological University, Houghton, MI 49931

Shashi P. Karna[*]
US Army Research Laboratory, Weapons and Materials Research Directorate,
ATTN: AMSRD-ARL-WM, Aberdeen Proving Ground, MD 21005-5069

**Abstract**

The electronic conduction of a novel, three-terminal molecular architecture, analogous to a heterojunction bipolar transistor is studied. In this architecture, two diode arms consisting of donor-acceptor molecular wires fuse through a ring, while a gate modulating wire is a π-conjugated wire. The calculated results show the enhancement or depletion mode of a transistor by applying a gate field along the positive or negative direction. A small gate field is required to switch on the current in the proposed architecture. The changes in the electronic conduction can be attributed to the intrinsic dipolar molecular architecture in terms of the evolution of molecular wavefunctions, specifically the one associated with the terphenyl group of the modulating wire in the presence of the gate field.

[*]*Corresponding Authors.*
Ravindra Pandey: *pandey@mtu.edu*; Tel: +1 (906) 487-2086, Fax: +1 (906) 487-2933;
Shashi P. Karna: *skarna@arl.army.mil*; Tel: +1 (410) 306-0723, Fax: +1 (410) 306-0723.

## 1. Introduction

In single molecules, electronic charge can be modulated by either electrical field or chemical effects, thereby opening up the possibility of their use as active elements in electronic devices. For example, a two-terminal molecular architecture consisting of a donor (D)-bridge (B)-acceptor (A) can show a significant rectification in current ($I$) under the applied bias voltage ($V$). Experimentally, many prototypes have been fabricated using single molecules in electronic devices [1-7] since the first theoretical study of a unimolecular diode by Aviram and Ratner [8].

Next, the technologically appealing feature will be a three-terminal molecular architecture with an offering of the possibility of switching current across the two arms and the external electric field modulation through the third arm of a molecule. Such architecture consisting of hybrid molecular diode has been fabricated [9], though its theory has not yet been established.

In the scientific literature, the examples of the three-terminal device (where the source-drain current is controlled by the gate voltage) include $C_{60}$ fullerene molecule [10, 11], carbon nanotubes [12, 13], organic self-assembled monolayer [14], organic molecules such as benzene-1,4-dithiolate [15, 16], and transition metal coordination complex [17].

In this paper, we report the results of a theoretical study of a novel, three-terminal architecture which is analogous to the heterojunction bipolar transistor. In this architecture, two diode arms consisting of donor-acceptor molecular wires fuse through a ring and a base or modulating wire is a π-conjugated wire. We note that it has recently been demonstrated that substitution by electron donating (D) or electron withdrawing (A) groups can effectively tune the electronic behavior of organic molecules and either lead to a simple shift of the electron transmission or to a more complex resonance effect [18]. In the present study, we explore the possible means of combining both D and A substitutions to control the functionality of the molecular transistor. The design of this unimolecular transistor further allows us to have a strong electrostatic coupling between the molecule and the gate, but a relatively weak coupling between the molecule and the source/drain electrode which is highly desirable in a three-terminal transistor architecture.

The diode characteristics of a DBA wire was focus of our recent study [19] where the intrinsic dipole nature of the molecule appears to determine its diode behavior. The larger the polarization of the system under an applied field, the more effective the electrical rectification is. The essence of the design of the proposed unimolecular transistor is to utilize the flexibility of molecular engineering to modify the electronic conducting channels (i.e. orbitals) of the molecular system and to magnify their response under an applied gate potential. There are three unique features in the proposed three-terminal architecture: (i) paring of *p*- and *n*-type functional units in a single molecule to introduce frontier orbitals for electron transport; (ii) unipolarity along the gate direction ensuring a strong coupling to the gate; and (iii) base molecule intervention in the evolution of molecular orbital wave characters to enhance electron transmission.

In the following section, we will give details of the three-terminal transistor and the method used for calculations of electrical characteristics of the device. The current ($I$) versus bias voltage ($V$) curves obtained under the gate voltage ($V_g$) are given in Sec. 3. The results are discussed in terms of the transmission functions in Sec. 4 and summary of the paper is given in Sec. 5.

## 2. Computational Details

The three-terminal molecular transistor consists of two diode units ABD and DBA where the donor (D) and acceptor (A) groups are 1,3-diaminobenzene and 1,2,4,5-tetracyanobenzene, respectively. A π-conjugated phenyl ring was used as the bridge (B) interconnected by a π-conjugated capacitor-a terphenyl unit (IC). It is constructed in line with the bipolar-junction architecture inherited from the microelectronic device designs of transistors. The molecule is symmetric along the source-drain direction (having a zero dipole component), while asymmetric in the gate direction (having a large dipole component). Such architecture is said to translate the internal conducting states with respect to the pseudo Fermi level of the system [20]. The application of a gate voltage ($V_g$) will significantly influence the current response and the electric current from source to drain is either enhanced or depleted depending on the direction of $V_g$.

The commonly used gold is taken to be source and drain electrodes. In order to eliminate the interfacial effects introduced by thiol groups, we used the C≡CH group as contacts to both termini. The molecule is considered to be coupled capacitively to the top and the bottom gates and the leakage current from the gates is taken to be negligibly small. The positive and negative gate fields are defined as the same and the opposite direction of the dipole moment of the entire molecule, respectively. For example, direction of a positive gate field pointing downwards is shown in Fig. 1.

The electron transport calculations were done in two steps. The first step was an electronic-structure calculation on the extended molecular complex instead of "an isolated molecule". It was performed in the framework of the density functional theory (DFT) with B3LYP functional form [21, 22] and the LanL2DZ basis set [23]. The extended molecular complex is the core scattering region of the molecular transport system, composed of the molecule and atomic-scale gold contacts in the form of atomic chains coupled to the source and drain electrodes. The geometrical configuration of the molecular complex obtained at the zero field was used for all the calculations considering that the electric filed employed does not distort the molecular configurations effectively.

An external electric field is applied in simulation of the applied gate potential. The correspondence between the electric field and the gate potential is listed in Table I considering the distance between the top and bottom gates to be around 32 Å. The Fock matrix $F$ and overlap matrix $S$ (corresponding to a non-orthogonal set of wavefunctions) from the self-consistent Kohn-Sham solution of the extended molecule electronic-structure calculation for each gate potential were then used for electron transport calculations based on the Green's function-based Landauer-Büttiker formalism [24-26]. Therefore, the Stark effect due to the electric field inside the molecule induced by the applied gate potential is included explicitly in the calculation.

The tunneling current ($I$) as a function of the applied bias ($V$) in such a device can be expressed as

$$I = \frac{2e}{h} \int_{-\infty}^{\infty} dE\, T(E,V)[f(E-\mu_1) - f(E-\mu_2)] \qquad (3)$$

where $\mu_1$ and $\mu_2$ are the electrochemical potentials in the source and drain electrodes under an external bias $V$, $f(E)$ is the Fermi-Dirac distribution function. $T(E,V)$ is the electron transmission function which can be calculated from a knowledge of the molecular energy levels and their coupling to the metallic contacts. Additional details of the calculations can be found elsewhere [19, 27, 28].

## 3. Results and Discussion

### 3.1 Tunneling Current - Enhancement/Depletion Gating Effect

Different from the conventional CMOS transistor, the unimolecular transistor has only one single molecule integrated between source, drain and gate. Due to the many body effects, the gate effect simulated by an external electric field in the presence of a source-drain potential is simply to increase the perturbation on the entire molecule. In such architecture, it is symmetric with respect to source and drain, but asymmetric in the gate direction. Thus, the perturbation due to the source-drain bias on the electronic structure is much smaller than the effect of the gate potential. In the low-bias regime considered, the change in the molecular spectra with respect to the source-drain bias is therefore ignored [29, 30].

The variation of the current as a function of the applied source-drain voltage is calculated at a gate field ranging from -9.25 to 9.25 ($10^8$ V/m). If we take the top-bottom gate distance to be around 32 Å, then the maximum gate potential ($V_g$) considered here is about 3 V. The calculated current ($I$) vs. drain voltage ($V_d$) curves under various gate potentials are plotted in Fig. 2. For a low bias (< 1.4 V) without applying a gate field, the current is found to be negligibly small. As the bias increases, however, the current tends to increase because of the resonant tunneling. An enhancement in current is observed for the positive gating as shown in Fig. 2, but depletion in current for the negative gating (not shown in Fig. 2).

Under positive gate fields, we find a gap and then a sudden increase in the $I$-$V_d$ curves. The current gets saturated when the $V_d$ is further increased. The observed Coulomb blockade is due to the weak coupling (no chemical bonding) between the molecule and the electrodes, where the molecule appears as an isolated quantum dot. The increment and saturation of current arising due to the resonant tunneling may be attributed to the specific molecular orbitals of the system which will be discussed more in detail in the following sections.

Increasing the gate electric field ($E_g$) from 0 to 9.25×$10^8$ V/m, one can observe switching of the molecule from a non-conducting state (OFF) to a conducting state (ON). The modulation of current (the ON/OFF ratio, defined as the current ratio with/without a gate field) also increases drastically with an increase in the gate potential ($V_g$), shown in the inset of Fig. 2. It reaches a maximum of about 70 for a gate of 6.17×$10^8$ V/m ($V_g \approx$ 2 V) and 160 for a gate of 9.25×$10^8$ V/m ($V_g \approx$ 3 V) at $V_d \approx$ 1.4 V. The gate electric field of 3.08×$10^8$ V/m ($V_g \approx$ 1 V), however, is not enough to switch on the current, general speaking. These values are significant for a transistor working in an enhancement mode. Under negative fields, the total current is reduced, suggesting a depletion working mode for the device.

## 3.2 Molecular Orbital Energy Shift – Polarization Effect

While maintaining a symmetric geometrical arrangement along the source-drain direction, the polarity along the gate direction, produces a permanent dipole moment along the $z$ direction, which ensures the strong electrostatic coupling between the molecule and the gate. And, in the current device architecture, polarizability of the molecule is enhanced by connecting two "p-n junctions" (i.e. the ABD molecules) tail to tail.

In Table I, we give the total energy and change in the electronic structure in terms of the dipole moment under a varying gate field. As a symbol of the polarization in the molecule along the applied gate potential, the dipole moment in this molecule varies significantly and almost linearly with the external electric field. The positive field enhances the dipole moment, and lowers the energy of the system, while the negative field reduces the dipole moment and increases the total energy.

The evolution of molecular orbital eigenvalue spectra with applied gate potential is plotted in Fig. 3. The molecular orbitals due to the donor (D) group on the HOMO side and the acceptor (A) group on the LUMO side are the frontier orbitals near the pseudo Fermi level $E_F$. Accordingly, the D-A orbital gap gets narrower with a positive field (enhancement mode) and opens up under a negative field (depletion mode). Note that the pure metallic orbitals from Au in the near-$E_F$ region (LUMO and LUMO+1, which are 2-fold degenerate) do not contribute to the conductance of the molecule.

In the low bias regime considered, contribution to molecule conductance mainly comes from the D orbitals (from the HOMO side), because the A orbitals are far above the Fermi level. Since the polarization effect induced by the positive gate potential moves the molecular energy levels upward (Table I), the molecular resonant states can be reached earlier relative to the cases when no or smaller gate potentials are applied. This partially accounts for the switching effect in the $I$-$V_d$ curve upon the application of a positive gate potential. Vice verse, the application of a negative gate potential moves the molecular energy levels downward, and therefore the molecular resonant states cannot be reached until a higher source-drain potential is applied. Thus a suppression of current is observed for all negative gate fields.

### 3.3 Evolution of Molecular Orbital Wave Characters – Base Molecule Intervention Effect

Analysis of the molecular orbitals reveals a significant evolution of the wavefunctions with the varying gate field in the range considered, as shown in Fig. 4. The IC (terphenyl) orbital near $E_F$ (highlighted in red in Fig. 3) lies below the four D highest occupied orbitals (labeled as D1, D2, D3, and D4 in sequence) at zero gate field. The IC orbital moves further down as a negative field increases in magnitude, but floats up when a positive field is applied, thereby mixing significantly with the D orbitals (especially the antisymmetric wavefunctions D1 and D3) at a field of $6.17 \times 10^8$ V/m. It lies above the second pair of D orbitals (D3 and D4) at a field strength of $9.25 \times 10^8$ V/m.

It is found that the IC orbital has a strong influence on the wave character of the D orbitals which determine the electron transmission property of the extended molecular system. In general, the lift-up of the IC orbital brings more terphenyl-ring component to the D wavefunctions. The consequence of this observation, however, is rather controversial. Note that the D-D connection is via one of the phenyl rings. Since the transmission of electrons requires the delocalization of the wavefunction, the bridging terphenyl ring's contribution to the corresponding D wavefunction is likely to be significant. Nevertheless, a strong terphenyl-ring character may otherwise work as an electron trap, making conduction of electrons difficult (see the small transmission peak in Fig. 5 which is associated with the IC orbital).

The transmission peaks due to the second pair of D orbitals (D3 and D4) make the most contribution to the electron conduction (Fig. 5), and correspond to the steep increase in current and thereafter the plateau of current (saturation). On the other hand, the peaks due to the first pair of D orbitals (D1 and D2) are rather small because of their association with the strong localization of the molecular wavefunctions.

A relatively strong transmission peak appears at -0.6 eV due to a large mixing between IC and D3 orbitals at a gate field of $6.17 \times 10^8$ V/m, suggesting that the terphenyl unit enclosed in the molecule acts more than just a passive capacitor electrostatically coupling the molecule to the gate. Its intrinsic character plays an active role in the enhancement gating of the three-terminal molecular transistor considered.

### 4. Summary

In summary, we have proposed a single-molecule model to realize the enhancement/ depletion mode of a three-terminal transistor by applying a gate field along the positive/negative direction. A small gate field is required to switch on the current, with the ON/OFF ratio reaching 160 at $V_g \approx 3$ V and $V_d \approx 1.4$ V. The asymmetric spatial structure leads to a sensitive coupling to electric field produced by the applied gate potential. The source-drain current is affected not only by the magnitude, but also by the direction, or the sign of the gate potential. One way enhances electric signal, but depletes the other. The evolution of wavefunctions, specifically the one associated with the terphenyl group in the presence of the gate-coupling molecular part as well as the shift in molecular orbital energies results in the enhancement of the source-drain current in the three-terminal transistor considered.

*Acknowledgement*-The work at Michigan Technological University was performed under support by the DARPA through contract number ARL-DAAD17-03-C-0115. The work at Army Research Laboratory (ARL) was supported by the DARPA MoleApps program and ARL-Director's Research Initiative-FY05-WMR01. Helpful discussions with S. Gowtham, K. C. Lau and R. Pati are acknowledged.

Table 1. The extended molecular complex under a series of applied gate fields: total energy, dipole moment, and energy of the HOMO level (due to the D group) with respect to the pseudo Fermi level $\Delta(\varepsilon_D - E_F)$. The distance between the top and bottom gates in the device architecture is taken to be 32 Å.

| Field ($10^8$ V/m) | Gate Potential[*] (V) | Energy (hartree) | Dipole (Debye) | $\Delta(\varepsilon_D - E_F)$ (eV) |
|---|---|---|---|---|
| -9.25 | -3 | -3733.596496 | 0.43 | -0.88 |
| -6.17 | -2 | -3733.596802 | 2.16 | -0.76 |
| -3.08 | -1 | -3733.597518 | 3.90 | -0.65 |
| 0.00 | 0 | -3733.598644 | 5.64 | -0.53 |
| 3.08 | 1 | -3733.600182 | 7.39 | -0.41 |
| 6.17 | 2 | -3733.602135 | 9.15 | -0.30 |
| 9.25 | 3 | -3733.604504 | 10.92 | -0.18 |

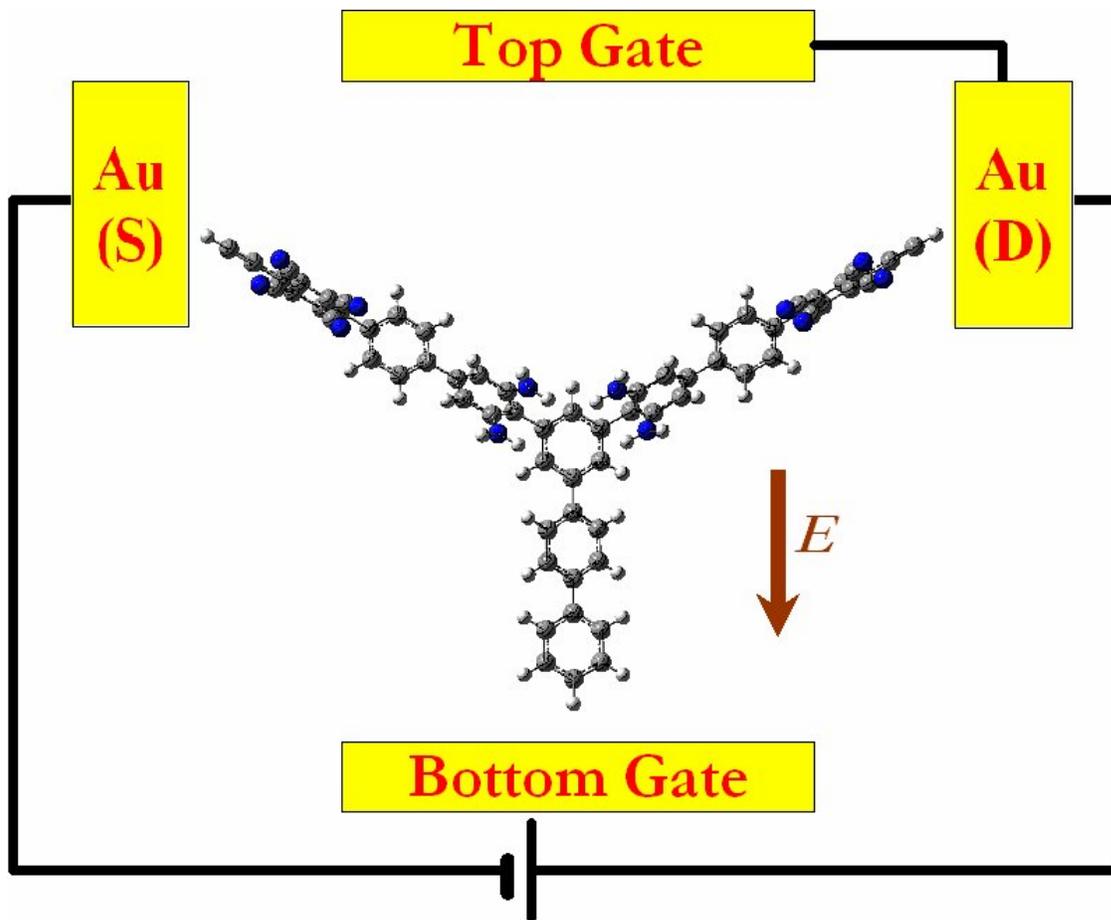

Fig. 1 Schematic illustration of the setup of a unimolecular transistor ABD-IC-DBA in measurement. C atoms are in gray spheres, N atoms in blue spheres, H atoms in small white spheres, and Au electrodes in yellow slabs. In this setup, a positive bias is applied upon the source (S) and the drain (D) termini, where the drain has a higher potential. A top-and-bottom double gate is applied generating a uniform positive gate field $E$ pointing downward in the same direction as that of the intrinsic dipole moment of the entire molecule.

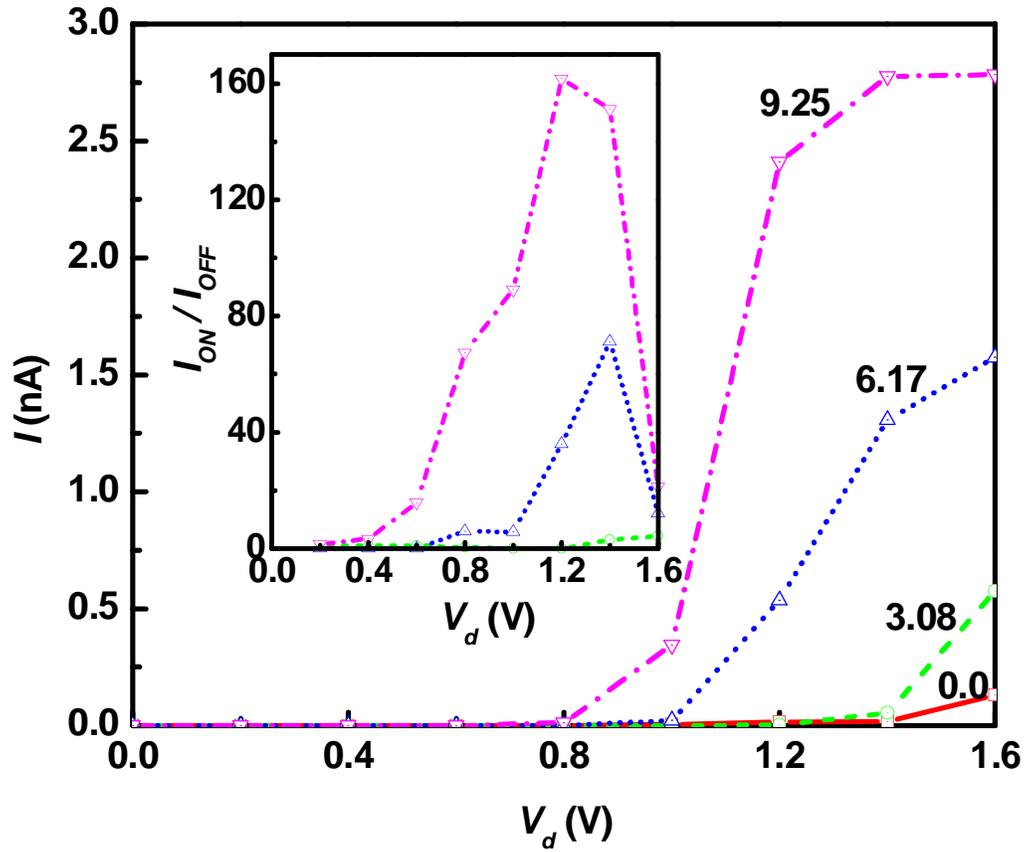

Fig. 2 $I$-$V_d$ curves for the molecular system under an applied positive gate field (labeled in the figure next to each curve in unit of $10^8$ V/m) in the range of 0.0 to 9.25×$10^8$ V/m. The calculated modulation factors (the current ratios $I_{ON}/I_{OFF}$) are also plotted in the inset for each field.

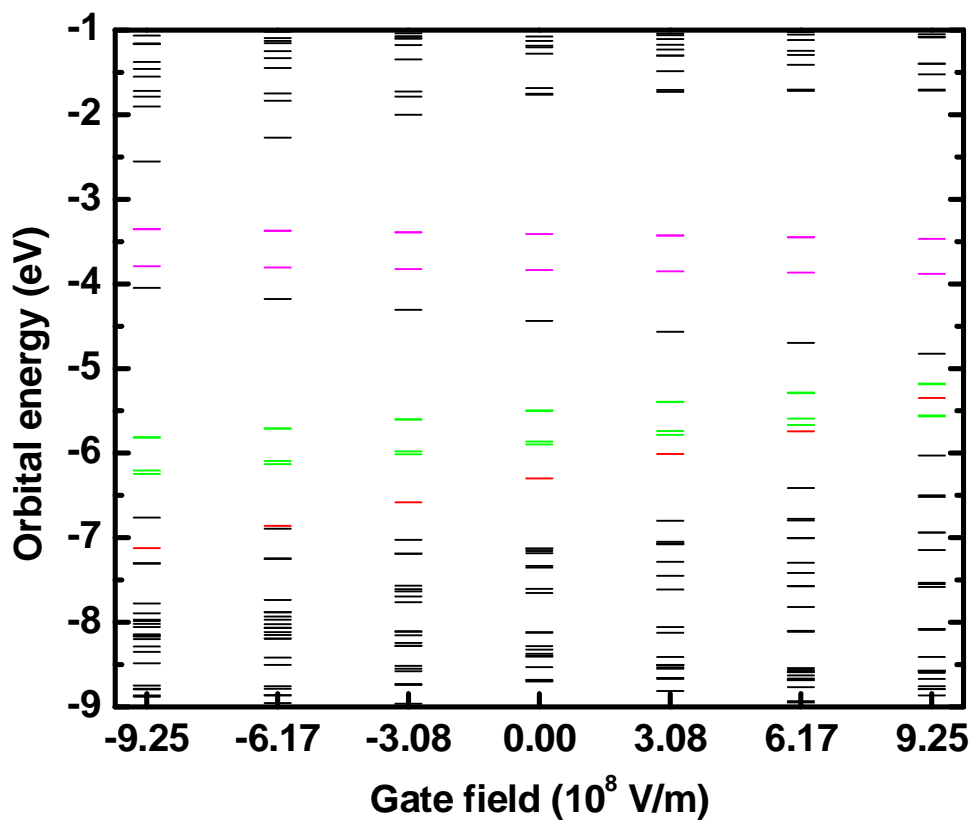

Fig. 3 Evolution of molecular spectra of the $Au_2$-ABD-IC-DBA-$Au_2$ extended molecule under a series of applied gate field. The highest occupied orbitals due to the donor group (D) are in green; the lowest unoccupied orbitals due to the acceptor group (A) are in magenta; the highest occupied orbital due to the IC unit is in red.

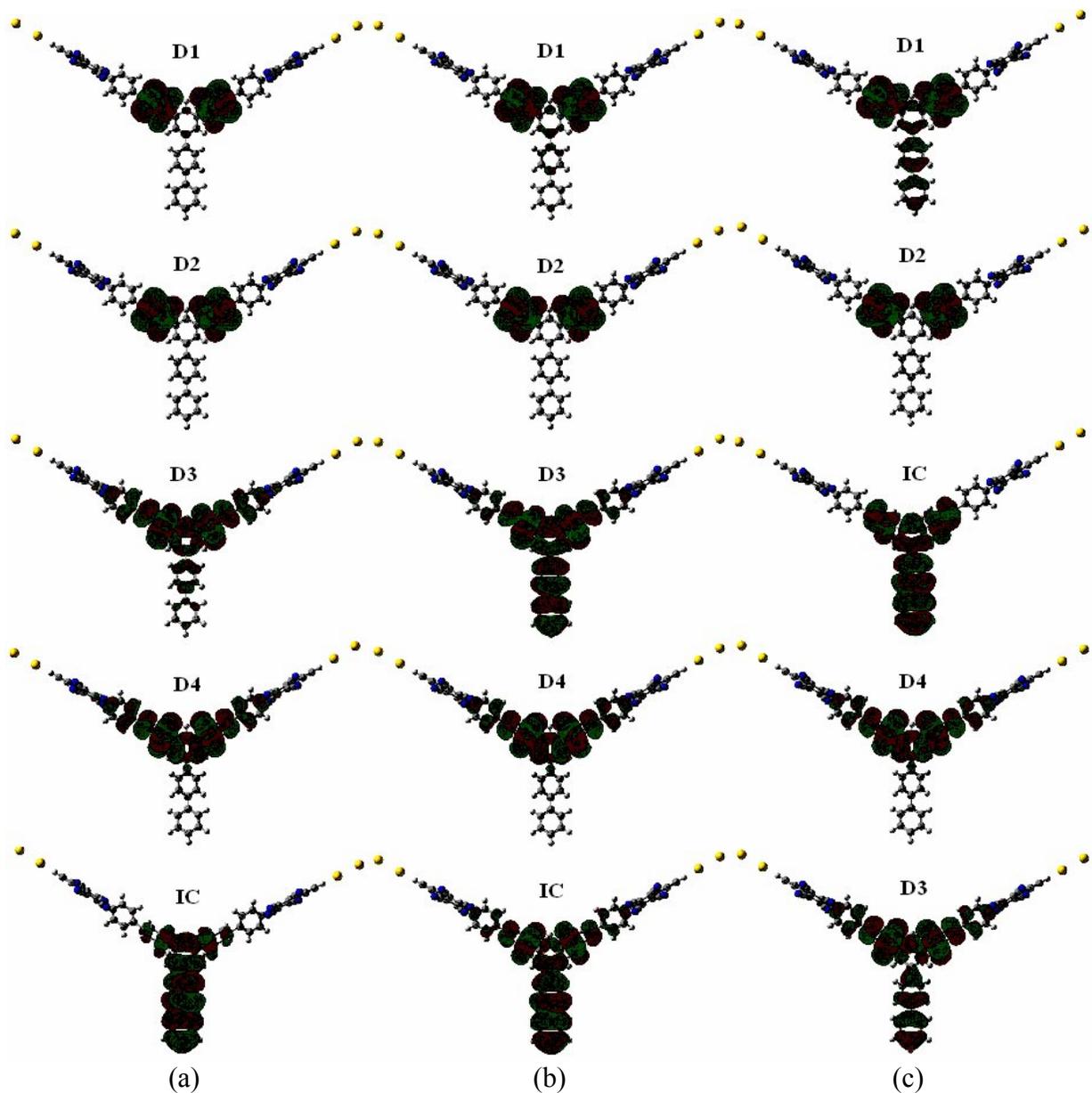

Fig. 4 Evolution of molecular orbitals of the $Au_2$-ABD-IC-DBA-$Au_2$ extended molecule under a series of applied gate field: (a) 0.0 V/m, (b) 6.17×10$^8$ V/m and (c) 9.25×10$^8$ V/m. The highest occupied orbitals mostly due to the donor group (D) at a zero field are labeled in sequence of D1, D2, D3 and D4. The highest occupied orbital due to the IC unit is labeled as IC.

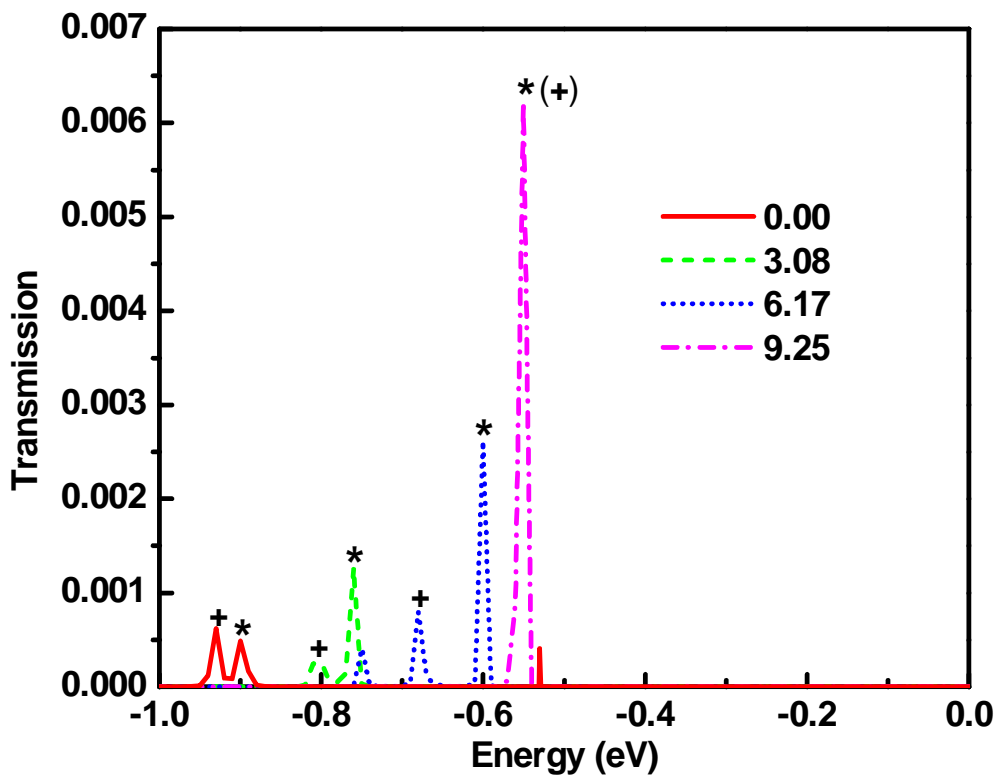

Fig. 5 Transmission functions for the ABD-IC-DBA molecular system under a series of applied gate field (labeled in the figure in unit of $10^8$ V/m). The transmission peaks due to the second pair of the highest occupied D orbitals D3 and D4 are labeled by * and +, respectively. The pseudo Fermi level is aligned to zero.